\documentclass[pra,aps,twocolumn,superscriptaddress]{revtex4}
\usepackage{hyperref}
\usepackage{bm}
\usepackage{epsfig}
\usepackage{amssymb}
\usepackage{graphicx}

\usepackage{amsmath}
\usepackage{epsfig}

\begin{document}
\title{Improving continuous-variable entanglement distribution by separable states}
\author{Ladislav Mi\v{s}ta, Jr.}
\affiliation{Department of Optics, Palack\' y University, 17.
listopadu 50,  772~07 Olomouc, Czech Republic}
\author{Natalia Korolkova}
\affiliation{School of Physics and Astronomy, University of St.
Andrews, North Haugh, St. Andrews, Fife, KY16 9SS, Scotland}

\date{\today}

\begin{abstract}
We investigate the physical mechanism behind the counterintuitive
phenomenon, the distribution of continuous-variable entanglement
between two distant modes by sending a third separable auxiliary
mode between them. For this purpose, we propose a new more simple
and more efficient protocol resulting in distributed entanglement
with more than an order of the magnitude higher logarithmic negativity than in the
previously proposed protocol. This new protocol shows that the
distributed entanglement originates from the entanglement of one
mode and the auxiliary mode used for distribution which is first
destroyed by local correlated noises and restored subsequently
by the interference of the auxiliary mode with the second distant
separable correlated mode.
\end{abstract}

\maketitle
\section{Introduction}

Quantum entanglement is a quintessence of quantum mechanics that
nowadays finds applications in the field of quantum communication
and information processing. Entanglement assisted communication
requires the communicating parties, Alice and Bob, to establish an
entangled state between their systems $a$ and $b$. This can be
done {\it only} by using a global operation, e.g. sending a
quantum system $c$ between them, as entanglement cannot be created
by local operations and classical communication (LOCC)
\cite{Werner_89}. This, however, does not necessarily imply that
the system $c$ must be entangled with the other two systems as
everybody would intuitively expect. Namely, for the quantum mixed
states only classical correlations and quantum interference
suffice to create entanglement \cite{Cubitt_03}. Specifically, for
mixed states one can establish a distillable entanglement between
two qubits $a$ and $b$ by using a third qubit $c$ being always
separable from the two-qubit system $(ab)$ \cite{Cubitt_03}. This
phenomenon has a number of implications for quantum communication,
which still have to be fully explored. For instance, quantum
teleportation \cite{ Bennett_93} does not require transmission of
an entangled quantum system between the sender and the receiver.

Distribution of entanglement without sending entanglement was
first suggested for qubits \cite{Cubitt_03} and only recently, an
alternative protocol for entanglement distribution by separable
states was proposed for Gaussian states of infinitely-dimensional
quantum systems and quantum variables with continuous
spectra-continuous variables (CVs) \cite{Mista_08}.

In this paper we study how the CV protocol works.
To that aim we propose a new protocol distributing entanglement with
more than an order of the magnitude higher logarithmic negativity
than the previous one \cite{Mista_08}. The proposed
protocol is more simple and allows to fully explain the physical mechanism underlying
the phenomenon of CV entanglement distribution by separable states.

The paper is organized as follows. Section~\ref{sec_1} contains a
brief introduction into the formalism of Gaussian states. In
Sec.~\ref{sec_2} we remind the original scheme for distribution of
CV entanglement by separable Gaussian states. Section \ref{sec_3}
is dedicated to the proposal of a new improved scheme for
establishing entanglement using separable ancilla. In
Sec.~\ref{sec_4} we explain the physics behind the distribution of
CV entanglement by separable states. Section~\ref{sec_5} covers
conclusion and comments on the experimental feasibility of the
protocol.
%%%%%%%%%%%%%%%%%%%%%%%%%%%%%%%%%%%%%%%%%%%%%%%%%%%%%%%%%%%%%%%%
\section{Gaussian states}\label{sec_1}

We consider three distinguishable infinite-dimensional quantum
systems $A,B$ and $C$, referred to as ``mode $A$,'' ``mode $B$,''
and ``mode $C$.'' The modes are described by three pairs of
canonically conjugate quadrature operators $x_{j},p_{j}$,
$j=A,B,C$ satisfying the canonical commutation rules
$[x_{j},x_{k}]=[p_{j},p_{k}]=0$, $[x_{j},p_{k}]=i\delta_{jk}$.
Defining the column vector
$\xi=(x_{A},p_{A},x_{B},p_{B},x_{C},p_{C})^{T}$ the
commutation rules can be written in the compact form
$[\xi_{j},\xi_{k}]=-i\Omega_{jk}$, where
%%%%%%%%%%%%%%%%%%%%%%%%%%%%%%%%%%%%%%%%%%%%%%%%%%%%%%%%%%%%%%%%%%%%%%%%%%%%%%%%%%%%%%%%%%%%%%%
\begin{eqnarray}\label{Omega}
\Omega=J\oplus J\oplus J,\quad J=\left(\begin{array}{cc}
0 & -1 \\
1 & 0 \\
\end{array}\right)
\end{eqnarray}
%%%%%%%%%%%%%%%%%%%%%%%%%%%%%%%%%%%%%%%%%%%%%%%%%%%%%%%%%%%%%%%%%%%%%%%%%%%%%%%%%%%%%%%%%%%%%%%%
is the standard three-mode symplectic matrix. All states of our
three-mode system can be represented in the real six-dimensional
phase space by the Wigner function \cite{Wigner_32} and Gaussian states are defined
as those states for which the Wigner function is Gaussian. Any
three-mode Gaussian state $\rho$ can be therefore fully
characterized by the real six-dimensional vector
$\langle\xi\rangle=\mbox{Tr}(\rho\xi)$ of phase-space displacements and by the
$6\times 6$ real symmetric covariance matrix (CM) $\gamma$ with
elements $\gamma_{jk}=\mbox{Tr}\left(\rho\{\xi_{j}-\langle\xi\rangle_{j}\openone,\xi_{k}-\langle\xi\rangle_{k}\openone\}\right)$,
$j,k=1,\ldots,6$, where $\{A,B\}\equiv AB+BA$. Due to the commutation rules the CM $\gamma$
satisfies the uncertainty relation \cite{Simon_87}
%%%%%%%%%%%%%%%%%%%%%%%%%%%%%%%%%%%%%%%%%%%%%%%%%%%%%%%%%%%%%%%%%%%%%%%%%%%%%%%%%%%%%%%%
\begin{equation}\label{uncertainty}
\gamma-i\Omega\geq0.
\end{equation}
%%%%%%%%%%%%%%%%%%%%%%%%%%%%%%%%%%%%%%%%%%%%%%%%%%%%%%%%%%%%%%%%%%%%%%%%%%%%%%%%%%%%%%%%%

In the following section we are interested in separability
properties of three-mode Gaussian states and therefore we need a
separability criterion for all possible $1\times 2$-mode
bipartitions, i.e. bipartitions  $A-(BC)$, $B-(AC)$ and $C-(AB)$. For
this purpose we can use the positive partial transpose (PPT)
criterion \cite{Peres_96,Horodecki_96} translated to Gaussian
states \cite{Duan_00,Simon_00,Werner_01,Giedke_01}. On the level
of CMs the partial transposition with respect to the mode $x$ transforms
the CM $\gamma$ to $\gamma^{(T_x)}=\Lambda_{x}\gamma\Lambda_{x}$
where $\Lambda_{x}$, $x=A,B,C$ are the diagonal matrices
$\Lambda_{A}=\sigma_{z}\oplus\openone\oplus\openone$,
$\Lambda_{B}=\openone\oplus\sigma_{z}\oplus\openone$,
$\Lambda_{C}=\openone\oplus\openone\oplus\sigma_{z}$, where
$\sigma_{z}$ is a Pauli diagonal matrix
$\sigma_{z}=\mbox{diag}(1,-1)$ and $\openone$ is a $2\times 2$
identity matrix. The PPT criterion then says that a three-mode
Gaussian state with CM $\gamma$ is separable with respect to
bipartition $x-(yz)$ (where $(x,y,z)$ is an even permutation of
$A,B,C$) if and only if the matrix $\gamma^{(T_x)}$ satisfies the
uncertainty relation (\ref{uncertainty}) \cite{Giedke_01}, i.e.
%%%%%%%%%%%%%%%%%%%%%%%%%%%%%%%%%%%%%%%%%%%%%%%%%%%%%%%%%%%%%%%%%%%%%%%%%%%%%%%%%%%%%%%%
\begin{equation}\label{criterion1}
\gamma^{(T_x)}-i\Omega\geq0.
\end{equation}
%%%%%%%%%%%%%%%%%%%%%%%%%%%%%%%%%%%%%%%%%%%%%%%%%%%%%%%%%%%%%%%%%%%%%%%%%%%%%%%%%%%%%%%%%
The PPT criterion can be expressed in another equivalent form that
can be sometimes easier to use in practice. It relies on the fact
that for any matrix $\gamma^{(T_x)}$ of a partially transposed state
there exists a symplectic matrix $S$, i.e. a real $6\times 6$
matrix satisfying the condition $S\Omega S^{T}=\Omega$, such
that $S\gamma^{(T_x)}S^{T}=\mbox{diag}(s_1,s_1,s_2,s_2,s_3,s_3)$
\cite{Williamson_36}. The nonnegative quantities $s_i$, $i=1,2,3$
are the so called symplectic eigenvalues of $\gamma^{(T_x)}$ and
$\gamma$ is separable with respect to the splitting $x-(yz)$ if and only if
$s_{i}\geq1$ $\forall i$ \cite{Vidal_02}. The symplectic eigenvalues can be
computed via the eigenvalues of the matrix $\Omega\gamma^{(T_x)}$ which are
$\left\{\pm is_1,\pm is_{2},\pm is_{3}\right\}$. For
complex three-mode states the symplectic eigenvalues are involved and it
is easier to use the PPT criterion formulated in terms of
the symplectic invariants \cite{Serafini_06}. The matrix $\gamma^{(T_x)}$ possesses
three such invariants denoted $I_1,I_2$ and $I_3=\mbox{det}(\gamma^{(T_{x})})$ that
can be calculated easily as coefficients of the characteristic
polynomial of the matrix $\Omega\gamma^{(T_{x})}$, i.e.
%%%%%%%%%%%%%%%%%%%%%%%%%%%%%%%%%%%%%%%%%%%%%%%%%%%%%%%%%%%%%%%%%
\begin{equation}\label{polynomial}
\mbox{det}(\Omega\gamma^{(T_{x})}-\mu\openone)=\mu^{6}+I_1\mu^{4}+I_2\mu^{2}+I_3.
\end{equation}
%%%%%%%%%%%%%%%%%%%%%%%%%%%%%%%%%%%%%%%%%%%%%%%%%%%%%%%%%%%%%%%%%
The criterion then says that for CM $\gamma$ mode $x$ is
separable from modes $(yz)$ if and only if
%%%%%%%%%%%%%%%%%%%%%%%%%%%%%%%%%%%%%%%%%%%%%%%%%%%%%%%%%%%%%%%%
\begin{equation}\label{Sigma}
\Sigma=\prod_{j=1}^{3}(s_{j}^2-1)=I_3-I_2+I_1-1\geq0
\end{equation}
%%%%%%%%%%%%%%%%%%%%%%%%%%%%%%%%%%%%%%%%%%%%%%%%%%%%%%%%%%%%%%%%%
holds. Strictly speaking, the criterion is not sufficient for
separability of states having some symplectic eigenvalue equal to
unity \cite{Serafini_06}. However, the sharp inequality $\Sigma>0$
is already sufficient for separability.
%%%%%%%%%%%%%%%%%%%%%%%%%%%%%%%%%%%%%%%%%%%%%%%%%%%%%%%%%%%%%%%%%%%%%%%%%%%%%%%%%%%%%%%%%
\section{The original  protocol}\label{sec_2}

%%%%%%%%%%%%%%%%%%%%%%%%%%%%%%%%%%%%%%%%%%%%%%%%%%%%%%%%%%%%%%%%%%%%%%%%%%%%%%%%%%%%%%%%%%%
\begin{figure}
%\centerline{\psfig{width=6.5cm,angle=0,file=fig1-NK.eps}}
\centerline{\psfig{width=7.5cm,angle=0,file=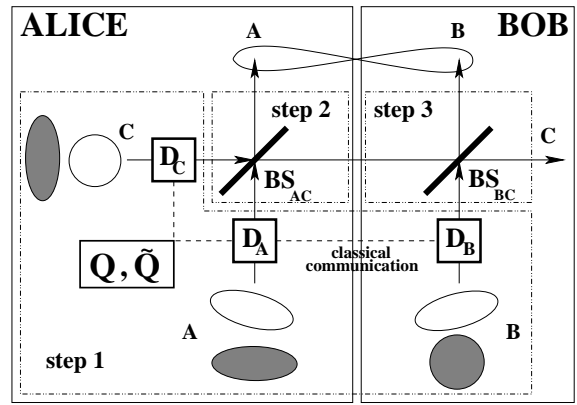}}
\caption{Two schemes of the protocol for distribution of CV
entanglement by separable Gaussian states. In the original scheme \cite{Mista_08}
(empty circle and ellipses) Alice's mode $A$ and Bob's distant mode $B$
are prepared in suitable rotated squeezed vacuum states while mode $C$ is
hold by Alice and it is in a vacuum state. In the improved scheme
of section \ref{sec_3} (hatched circle and ellipses) modes $A$ and $C$ hold by Alice are
in the momentum and position squeezed vacuum states, respectively, and Bob's mode $B$
is in a vacuum state. All the three modes are then displaced by displacements
$D_{A},D_{B}$ and $D_{C}$ distributed randomly with Gaussian distribution
with covariance matrix $Q$ ($\tilde{Q}$ for the improved protocol) after
which the modes are in a fully separable state (step 1). Mixing of modes $A$ and $C$ on a
balanced beam splitter $BS_{AC}$ entangles mode $A$ with the pair
of modes $(BC)$ while mode $B$ is separable from $(AC)$ and mode $C$
is separable from $(AB)$ (step 2). Mixing of modes $B$ and $C$ on a balanced
beam splitter $BS_{BC}$ finally entangles $A$ and $B$ wile $C$
still remains separable from $(AB)$ (step 3). See text for details.}
\label{fig1}
\end{figure}
%%%%%%%%%%%%%%%%%%%%%%%%%%%%%%%%%%%%%%%%%%%%%%%%%%%%%%%%%%%%%%%%%%%%%%%%%%%%%%%%%%%%%%%%%%%

The original protocol for distribution
of CV entanglement by separable Gaussian states \cite{Mista_08} is
depicted in Fig.~\ref{fig1} (see figure caption for details). For this protocol the contours of
the phase-space distributions of the input modes are shown as
empty circle and ellipses. The essential ingredient of the protocol is  a
fully separable three-mode Gaussian state $\rho_{ABC}$ that
transforms appropriately under beam splitter operations on
certain pairs of modes. More precisely, $\rho_{ABC}$ transforms under
$BS_{AC}$ operation on modes $A$
and $C$ onto a state separable with respect to two bipartitions
$B-(AC)$ and $C-(AB)$ (two-mode biseparable state \cite{Giedke_01}). This
state further transforms under $BS_{BC}$ operation on modes $B$ and
$C$ onto the state separable with respect to only one bipartition
$C-(AB)$ (one-mode biseparable state
\cite{Giedke_01}). The construction of such a state is therefore
the key nontrivial task that has to be solved in order to find the
protocol. The requirements of the protocol are counterintuitive:
by using LOCC and sending of a separable ancilla from Alice to Bob
to create entanglement between them. As no entanglement is shared
beforehand or sent by Alice to Bob, one can ask a natural
question where the entanglement in the protocol comes from.
In this original protocol,
there is no a straightforward simple explanation for the emergence of entanglement, in particular due to the complex structure of the quantum states involved.
For instance, the CMs occurring in the protocol possess correlations between position and
momentum quadratures and it is fairly possible that this kind of correlations is responsible for
the phenomenon. Similarly, it is not obvious which role is played by the squeezed state on
Bob's side and whether it has necessarily to be squeezed for the entanglement distribution to occur.

In the following section we show that neither of these properties accounts for the phenomenon.
We construct a more simple protocol where all the participating states have zero correlations
between position and momentum quadratures and where Bob's mode is initially in the vacuum state
(which is a completely classical state). Further, the new protocol exhibits a substantially better
performance than the original one as it establishes entanglement between modes $A$ and $B$ with
more than an order of the magnitude higher logarithmic negativity. Finally, due to its simplicity the protocol gives a clear physical
explanation of the counterintuitive phenomenon of entanglement distribution by separable states.

%%%%%%%%%%%%%%%%%%%%%%%%%%%%%%%%%%%%%%%%%%%%%%%%%%%%%%%%%%%%%%%%%%%%%%%%%%%%%%%%%%%%%%%%%%%%%%%%%%%%%%%%%%
\section{The improved protocol}\label{sec_3}

In contrast to the original protocol \cite{Mista_08} where the fully separable state of step 1 was constructed first,
we start with designing the state in step 2. That is, by local operations on a pair
of modes $(AC)$ and mode $B$ and classical communication we construct a three-mode mixed Gaussian state
in which mode $A$ is entangled with modes $(BC)$, mode $C$ is separable from modes $(AB)$ and mode $B$ is separable
from modes $(AC)$. Initially, we assume mode $B$ in the vacuum state and modes $A$ and $C$ in the two-mode squeezed
vacuum state. The corresponding CMs read as
%%%%%%%%%%%%%%%%%%%%%%%%%%%%%%%%%%%%%%%%%%%%%%%%%%%%%%%%%%%%%%%%%%%%%%%%%%%%%%%%%%%%%%%%%
%%%%%%%%%%%%%%%%%%%%%%%%%%%%%%%%%%%%%%%%%%%%%%%%%%%%%%%%%%%%%%%%%%%%%%%%%%%%%%%%%%%%%%%%%%%%%%
\begin{eqnarray}\label{gammaACgammaB}
\gamma_{0,AC}=\left(\begin{array}{cc}
\cosh(2t)\openone & \sinh(2t)\sigma_{z} \\
\sinh(2t)\sigma_{z} & \cosh(2t)\openone\\
\end{array}\right),\quad \gamma_{0,B}=\openone,
\end{eqnarray}
%%%%%%%%%%%%%%%%%%%%%%%%%%%%%%%%%%%%%%%%%%%%%%%%%%%%%%%%%%%%%%%%%%%%%%%%%%%%%%%%%%%%%%%%%%%%%
where $t\geq0$ is the squeezing parameter. The matrix $\gamma_{0,AC}^{(T_{C})}$ has
 the lower symplectic eigenvalue equal to $\nu_{\rm in}=e^{-2t}<1$ for $t>0$ and thus modes
 $A$ and $C$ are entangled. Now we take the CM of the entire
 three-mode system,
%%%%%%%%%%%%%%%%%%%%%%%%%%%%%%%%%%%%%%%%%%%%%%%%%%%%%%%%%%%%%%%%%%%%%%%%%%%%%%%%%%%%%%%%%%%%%%
\begin{eqnarray}\label{gammaABC}
\gamma_{ABC}=\left(\begin{array}{ccc}
\cosh(2t)\openone & 0 & \sinh(2t)\sigma_{z} \\
0 & \openone & 0 \\
\sinh(2t)\sigma_{z} & 0 & \cosh(2t)\openone\\
\end{array}\right),
\end{eqnarray}
%%%%%%%%%%%%%%%%%%%%%%%%%%%%%%%%%%%%%%%%%%%%%%%%%%%%%%%%%%%%%%%%%%%%%%%%%%%%%%%%%%%%%%%%%%%%%
 and we add to it a nonnegative multiple of a suitable positive noise matrix
 $P$ such that the state described by CM
 %%%%%%%%%%%%%%%%%%%%%%%%%%%%%%%%%%%%%%%%%%%%%%%%%%%%%%%%%%%%%%%%%%%%%%%%%%%%%%%%%%%%%%%%%%%
 \begin{equation}\label{tildegamma2a}
 \gamma_{2}=\gamma_{ABC}+xP, \qquad x\geq0,
 \end{equation}
 %%%%%%%%%%%%%%%%%%%%%%%%%%%%%%%%%%%%%%%%%%%%%%%%%%%%%%%%%%%%%%%%%%%%%%%%%%%%%%%%%%%%%%%%%%%%
 possesses entanglement between mode $A$ and modes $(BC)$ while it is separable
 with respect to the other two bipartitions. The matrix $P$ can be found using the method
 developed in \cite{Giedke_01} for construction of various three-mode entangled Gaussian states.
 It utilizes a two-mode version of the separability criterion represented by Eq.~(\ref{criterion1})
 according to which a two-mode CM $\gamma_{0,AC}$ is entangled if and only if the matrix
$\gamma_{0,AC}-iJ\oplus(-J)$ has a negative eigenvalue. By adding to the CM a sufficiently large
positive multiple of sum of the projectors onto the subspaces spanned by real and imaginary parts of the eigenvector
corresponding to the negative eigenvalue we can destroy the entanglement between modes $A$ and $C$. Extending
now the projectors suitably and adding a multiple of sum of the extensions to the CM $\gamma_{ABC}$ we can finally
construct a state with desired separability properties.
 The negative eigenvalue is easy to find and it reads as $\lambda=-(1-e^{-2t})<0$ for $t>0$ and the corresponding eigenvector is
 $p_{\lambda}=p_{1}+ip_{2}$, where $p_{1}=(0,-1,0,-1)^{T}$ and
 $p_{2}=(1,0,-1,0)^{T}$. The sought noise matrix then can be constructed from the following $6\times1$
 extensions of the vectors $p_{1}$ and $p_{2}$:
%%%%%%%%%%%%%%%%%%%%%%%%%%%%%%%%%%%%%%%%%%%%%%%%%%%%%%%%%%%%%%%%%%%%%%%%%%%%%%%%%%%%%%%%%
\begin{eqnarray}\label{tildeq}
q_{1}&=&(0,-1,0,2,0,-1)^{T},\nonumber\\
q_{2}&=&(1,0,2,0,-1,0)^{T},
\end{eqnarray}
%%%%%%%%%%%%%%%%%%%%%%%%%%%%%%%%%%%%%%%%%%%%%%%%%%%%%%%%%%%%%%%%%%%%%%%%%%%%%%%%%%%%%%%%%
as $P=q_{1}q_{1}^{T}+q_{2}q_{2}^{T}$. The matrix
is by construction positive and the addition of its sufficiently large nonnegative multiple $xP$
to the CM (\ref{gammaABC}) smears the initial entanglement between modes $A$ and $C$
such that $A$ is entangled with $(BC)$ yet $C$ is separable from $(AB)$ and, furthermore, $B$ is separable from $(AC)$.
The separability of mode $B$ from the pair of modes $(AC)$ is obvious as the CM (\ref{tildegamma2a}) can be created
from the product state with CM $\gamma_{0,AC}\oplus\gamma_{0,B}$ by local random correlated displacements distributed
with Gaussian distribution described by the correlation matrix $xP$. The threshold value $x_{\rm sep}$
of $x$ such that for $x\geq x_{\rm sep}$ the state with CM (\ref{tildegamma2a}) is separable with respect
to $C-(AB)$ splitting can be found from the requirement of positivity of the partial transpose of the
state described by CM (\ref{tildegamma2a}) with respect to mode $C$ that is sufficient
for separability of the $1\times2$ system \cite{Werner_01}. The state has a positive partial transpose
if and only if the symplectic eigenvalues $\tau_{j}$, $j=1,2,3$,
of the matrix $\gamma_{2}^{(T_C)}$ satisfy $\tau_{j}\geq1$ for all $j$ \cite{Vidal_02}.
The lowest symplectic eigenvalue reads
%%%%%%%%%%%%%%%%%%%%%%%%%%%%%%%%%%%%%%%%%%%%%%%%%%%%%%%%%%%%%%%%%%%%%%%%%%%%%%%%%%%%%%%%%%%%%%%%%%%%%%%%%%%%%%%%%%%%
\begin{equation}\label{nu}
\tau_{3}=\frac{\sqrt{\left(1+6x+e^{-2t}\right)^{2}-32x^2}-\left(1+2x-e^{-2t}\right)}{2}
\end{equation}
%%%%%%%%%%%%%%%%%%%%%%%%%%%%%%%%%%%%%%%%%%%%%%%%%%%%%%%%%%%%%%%%%%%%%%%%%%%%%%%%%%%%%%%%%%%%%%%%%%%%%%%%%%%%%%%%%%%%
and the threshold value $x_{\rm sep}$ can be calculated from the condition $\tau_{3}=1$. After
some algebra one can show that the CM (\ref{tildegamma2a})
represents a state that is separable with respect to $C-(AB)$
splitting if $x\geq x_{\rm sep}$, where
%%%%%%%%%%%%%%%%%%%%%%%%%%%%%%%%%%%%%%%%%%%%%%%%%%%%%%%%%%%%%%%%%%%%%%%%%%%%%%%%%%%%%%%%%%%%%%%%%%%%%%%%%%%%%%%%%%%%
\begin{equation}\label{tildexsep}
x_{\rm sep}=\frac{e^{2t}-1}{2}.
\end{equation}
%%%%%%%%%%%%%%%%%%%%%%%%%%%%%%%%%%%%%%%%%%%%%%%%%%%%%%%%%%%%%%%%%%%%%%%%%%%%%%%%%%%%%%%%%%%%%%%%%%%%%%%%%%%%%%%%%%%%
Entanglement of mode $A$ with a pair of modes $(BC)$ can be verified along the same lines. The lowest symplectic
eigenvalue of the matrix $\gamma_{2}^{(T_{A})}$ attains the form:
%%%%%%%%%%%%%%%%%%%%%%%%%%%%%%%%%%%%%%%%%%%%%%%%%%%%%%%%%%%%%%%%%%%%%%%%%%%%%%%%%%%%%%%%%%%%%%%%%%%%%%%%%%%%%%%%%%%%
\begin{equation}\label{omega}
\omega_{3}=\frac{1+6x+e^{-2t}-\sqrt{\left(1+2x-e^{-2t}\right)^{2}+32x^2}}{2}
\end{equation}
%%%%%%%%%%%%%%%%%%%%%%%%%%%%%%%%%%%%%%%%%%%%%%%%%%%%%%%%%%%%%%%%%%%%%%%%%%%%%%%%%%%%%%%%%%%%%%%%%%%%%%%%%%%%%%%%%%%%
and for $x>0$ and $t>0$ it satisfies the inequality $\omega_{3}<1$ which proves the entanglement between $A$ and $(BC)$.
Note that this entanglement is essential for the performance of the studied protocol as the second beam splitter $BS_{BC}$
alone cannot entangle mode $A$ with mode $B$ without $A$ being entangled with $(BC)$.

Having found the desired CM of step 2 we can return back to step 1 by implementing the inverse
of the balanced $BS_{AC}$ on modes $A$ and $C$. The beam splitter is described by the matrix
%%%%%%%%%%%%%%%%%%%%%%%%%%%%%%%%%%%%%%%%%%%%%%%%%%%%%%%%%%%%%%%%%%%%%%%%%%%%%%%%%%%%%%%%%%%%%%
\begin{eqnarray}\label{UAC}
U_{AC}=\left(\begin{array}{ccc}
\frac{1}{\sqrt{2}}\openone & 0 & \frac{1}{\sqrt{2}}\openone \\
0 & \openone & 0\\
\frac{1}{\sqrt{2}}\openone& 0 & -\frac{1}{\sqrt{2}}\openone \\
\end{array}\right),
\end{eqnarray}
%%%%%%%%%%%%%%%%%%%%%%%%%%%%%%%%%%%%%%%%%%%%%%%%%%%%%%%%%%%%%%%%%%%%%%%%%%%%%%%%%%%%%%%%%%%%
and the inverse operation $U_{AC}^{-1}=U_{AC}^{T}=U_{AC}$ transforms
the CM (\ref{tildegamma2a}) as
%%%%%%%%%%%%%%%%%%%%%%%%%%%%%%%%%%%%%%%%%%%%%%%%%%%%%%%%%%%%%%%%%%%%%%%%%%%%%%%%%%%%%%%%%%
\begin{equation}\label{tildegamma1}
\gamma_{1}=U_{AC}^{T}\gamma_{2}U_{AC}=\gamma_{A}\oplus\gamma_{B}\oplus\gamma_{C}+\tilde{Q},
\end{equation}
%%%%%%%%%%%%%%%%%%%%%%%%%%%%%%%%%%%%%%%%%%%%%%%%%%%%%%%%%%%%%%%%%%%%%%%%%%%%%%%%%%%%%%%%%%
where the CM $\gamma_{B}=\gamma_{0,B}=\openone$ is the vacuum CM and
%%%%%%%%%%%%%%%%%%%%%%%%%%%%%%%%%%%%%%%%%%%%%%%%%%%%%%%%%%%%%%%%%%%%%%%%%%%%%%%%%%%%%%%%%%
\begin{eqnarray}\label{product}
\gamma_{A,C}=\left(\begin{array}{cc}
e^{\pm2t} & 0 \\
0 & e^{\mp2t} \\
\end{array}\right)
\end{eqnarray}
%%%%%%%%%%%%%%%%%%%%%%%%%%%%%%%%%%%%%%%%%%%%%%%%%%%%%%%%%%%%%%%%%%%%%%%%%%%%%%%%%%%%%%%%%%
are CMs of the momentum and position squeezed vacuum states, respectively. The structure of CM (\ref{tildegamma1})
reveals that the state of step 1 can be created from the product of two orthogonally squeezed states
and a vacuum state by local random displacements distributed according to the Gaussian distribution with
correlation matrix $\tilde{Q}=xU_{AC}^{T}PU_{AC}$ that has the following form:
%%%%%%%%%%%%%%%%%%%%%%%%%%%%%%%%%%%%%%%%%%%%%%%%%%%%%%%%%%%%%%%%%%%%%%%%%%%%%%%%%%%%%%%%%%%%%%
\begin{eqnarray}\label{tildeQ}
\tilde{Q}=x\left(\begin{array}{ccc}
\openone-\sigma_{z} & \sqrt{2}(\sigma_{z}-\openone) & 0 \\
\sqrt{2}(\sigma_{z}-\openone) & 4\cdot\openone & \sqrt{2}(\sigma_{z}+\openone) \\
0 & \sqrt{2}(\sigma_{z}+\openone) & \sigma_{z}+\openone\\
\end{array}\right).
\end{eqnarray}
%%%%%%%%%%%%%%%%%%%%%%%%%%%%%%%%%%%%%%%%%%%%%%%%%%%%%%%%%%%%%%%%%%%%%%%%%%%%%%%%%%%%%%%%%%%%%
This implies immediately that the state is fully separable as is the basic requirement of our protocol.

The improved protocol for
distribution of CV entanglement by separable
Gaussian states is depicted in Fig.~\ref{fig1}. The phase-space
contours of the input modes for this protocol are represented by
hatched circle and ellipses. As follows from the CM
$\gamma_{1}$ designed in Eqs. (\ref{tildegamma1}) and
(\ref{product}), the input modes $A$ and $C$ are both in a squeezed
state with reduced noise in two orthogonal quadratures
respectively and mode $B$ is in a vacuum state. In the step 1
Alice holding a pair of modes $(AC)$ and Bob holding mode $B$
prepare by LOCC a Gaussian state characterized by the CM
(\ref{tildegamma1}). In the next step 2, Alice superimposes mode
$A$ and $C$ on a balanced beam splitter $BS_{AC}$. Thus she
creates a state with CM (\ref{tildegamma2a}) that reads explicitly
%%%%%%%%%%%%%%%%%%%%%%%%%%%%%%%%%%%%%%%%%%%%%%%%%%%%%%%%%%%%%%%%%%%
\begin{eqnarray}\label{tildegamma2b}
\gamma_{2}=\left(\begin{array}{ccc}
a\openone & 2x\sigma_{z} & b\sigma_{z} \\
2x\sigma_{z} & (1+4x)\openone & -2x\openone \\
b\sigma_{z} & -2x\openone & a\openone\\
\end{array}\right),
\end{eqnarray}
%%%%%%%%%%%%%%%%%%%%%%%%%%%%%%%%%%%%%%%%%%%%%%%%%%%%%%%%%%%%%%%%%%%%%%%%%%%%%%%%%%%%%%%%%%%%%
where $a=\cosh(2t)+x$ and $b=\sinh(2t)-x$. As was shown above, the state is separable with
respect to $B-(AC)$ splitting and for $x\geq x_{\rm sep}$, where $x_{\rm sep}$ is
given in Eq.~(\ref{tildexsep}), also with respect to $C-(AB)$ splitting.

The protocol is finalized by step 3 in which Bob mixes the mode $C$ received from Alice
with his mode $B$ on another balanced beam splitter $BS_{BC}$ described by the matrix
%%%%%%%%%%%%%%%%%%%%%%%%%%%%%%%%%%%%%%%%%%%%%%%%%%%%%%%%%%%%%%%%%%%
\begin{eqnarray}\label{UBC}
U_{BC}=\left(\begin{array}{ccc}
\openone & 0 & 0\\
0 & \frac{1}{\sqrt{2}}\openone &  \frac{1}{\sqrt{2}}\openone\\
0 & \frac{1}{\sqrt{2}}\openone&  -\frac{1}{\sqrt{2}}\openone\\
\end{array}\right).
\end{eqnarray}
%%%%%%%%%%%%%%%%%%%%%%%%%%%%%%%%%%%%%%%%%%%%%%%%%%%%%%%%%%%%%%%%%%%%%
The beam splitter transforms
the CM (\ref{tildegamma2b}) to CM $\gamma_{3}=U_{BC}\gamma_{2}U_{BC}^{T}$ that reads explicitly as
%%%%%%%%%%%%%%%%%%%%%%%%%%%%%%%%%%%%%%%%%%%%%%%%%%%%%%%%%%%%%%%%%%%%%%%%%%%%%%%%%%%%%%%%
\begin{eqnarray}\label{tildegamma3}
\gamma_{3}=\left(\begin{array}{ccc}
a\openone & \frac{2x+b}{\sqrt{2}}\sigma_{z} & \frac{2x-b}{\sqrt{2}}\sigma_{z} \\
\frac{2x+b}{\sqrt{2}}\sigma_{z} & \frac{1+a}{2}\openone & \frac{1+4x-a}{2}\openone \\
\frac{2x-b}{\sqrt{2}}\sigma_{z} & \frac{1+4x-a}{2}\openone & \frac{1+8x+a}{2}\openone\\
\end{array}\right),
\end{eqnarray}
%%%%%%%%%%%%%%%%%%%%%%%%%%%%%%%%%%%%%%%%%%%%%%%%%%%%%%%%%%%%%%%%%%%%%%%%%%%%%%%%%%%%%%%%
where $a$ and $b$ are defined as in Eq.~(\ref{tildegamma2b}). The interference on the beam splitter $BS_{BC}$
finally entangles modes $A$ and $B$. This can be seen by computing the lower symplectic eigenvalue
$\nu$ of the matrix $\gamma_{3,AB}^{(T_{B})}$ corresponding to the reduced state of modes $A$ and $B$.
We express the CM $\gamma_{3,AB}$ in the block form
%%%%%%%%%%%%%%%%%%%%%%%%%%%%%%%%%%%%%%%%%%%%%%%%%%%%%%%%%%%%%%%%%%%%%%%%%%%%%%%%%%%%%%%%%%%%%%%
\begin{eqnarray}\label{block}
\gamma_{3,AB}=\left(\begin{array}{cc}
A & C \\
C^{T} & B \\
\end{array}\right),
\end{eqnarray}
%%%%%%%%%%%%%%%%%%%%%%%%%%%%%%%%%%%%%%%%%%%%%%%%%%%%%%%%%%%%%%%%%%%%%%%%%%%%%%%%%%%%%%%%%%%%%%%%
where $A,B$ and $C$ are $2\times2$ submatrices.  The eigenvalue then reads as \cite{Vidal_02}
%%%%%%%%%%%%%%%%%%%%%%%%%%%%%%%%%%%%%%%%%%%%%%%%%%%%%%%%%%%%%%%%%%%%%%%%%%%%%%%%%%%%%%%%%%%%%%%
\begin{eqnarray}\label{ABsymplectic}
\nu=\sqrt{\frac{\kappa-\sqrt{\kappa^2-4\mbox{det}(\gamma_{3,AB})}}{2}},
\end{eqnarray}
%%%%%%%%%%%%%%%%%%%%%%%%%%%%%%%%%%%%%%%%%%%%%%%%%%%%%%%%%%%%%%%%%%%%%%%%%%%%%%%%%%%%%%%%%%%%%%%%
where
%%%%%%%%%%%%%%%%%%%%%%%%%%%%%%%%%%%%%%%%%%%%%%%%%%%%%%%%%%%%%%%%%%%%%%%%%%%%%%%%%%%%%%%%%%%%%%%%%%%
\begin{eqnarray}\label{ABinvariants}
\mbox{det}\left(\gamma_{3,AB}\right)&=&\left[\frac{1+\cosh\left(2t\right)}{2}+\left(e^{-2t}+\frac{1}{2}\right)x\right]^{2},\nonumber\\
\kappa&=&\mbox{det}A+\mbox{det}B-2\mbox{det}C\nonumber\\
&=&a^2+\left(b+2x\right)^2+\frac{\left(a+1\right)^2}{4}
\end{eqnarray}
%%%%%%%%%%%%%%%%%%%%%%%%%%%%%%%%%%%%%%%%%%%%%%%%%%%%%%%%%%%%%%%%%%%%%%%%%%%%%%%%%%%%%%%%%%%%%%%%%%%
and it can be used to quantify the amount of distributed
entanglement by calculating the logarithmic negativity given by
the formula $E_{\mathcal{N}}=-\log_{2}\nu$ \cite{Vidal_02}. For
$t>0$ and $x=x_{\rm sep}=(e^{2t}-1)/2$ we get $\nu<1$ and
therefore modes $A$ and $B$ are entangled for an arbitrarily small
nonzero squeezing. Numerical analysis further reveals that the
symplectic eigenvalue decreases monotonously with increasing
squeezing and approaches the value of $\nu=1/3$ corresponding to
$E_{\mathcal{N}}\approx1.585$ ebits in the limit of infinite
squeezing  while all the separability properties of the states
involved remain preserved. To illustrate this, we have calculated
the output entanglement for some realistic values of squeezing
using Eqs.~(\ref{ABsymplectic}) and (\ref{ABinvariants}). Thus
$e^{2t}=2$ corresponding to 3dB of input squeezing results in the
symplectic eigenvalue of $\nu\approx0.6589$ which gives the
logarithmic negativity of $E_{\cal N}\approx0.6019$ ebits. For
$e^{2t}=10$ corresponding to 10dB squeezing that was recently
achieved experimentally \cite{Schnabel_08} we get
$\nu\approx0.3968$ and $E_{\cal N}\approx1.3334$ ebits.

It is interesting to compare the latter logarithmic negativity with
the maximum logarithmic negativity that can be obtained with
the original protocol \cite{Mista_08}. Numerical optimization of the original
protocol with respect to the noise parameter $x$ and the initial
squeezings (denoted by $d\pm r$ in \cite{Mista_08}) reveals that
the minimal lower symplectic eigenvalue of the partially
transposed final entangled state distributed between Alice and Bob
reads as $\nu\approx0.9299$ and the corresponding logarithmic
negativity is $E_{\cal N}\approx0.1048$ ebits.
Thus the logarithmic negativity of the entangled state created
between distant locations in the improved protocol is more than an order of
the magnitude higher than in the original protocol. Furthermore, also in this
improved protocol the mode $C$ remains separable from modes $(AB)$
after the step 3. This can be shown using the separability criterion based on symplectic
invariants \cite{Serafini_06}. For example, for $e^{2t}=2$ the criterion gives
$\Sigma=1>0$ which implies separability of mode $C$ from $(AB)$.
%%%%%%%%%%%%%%%%%%%%%%%%%%%%%%%%%%%%%%%%%%%%%%%%%%%%%%%%%%%%%%%%%%%%%%%%%%%%%%%%%%%%%%%%%%%%%%%
\section{Explanation of the protocol}\label{sec_4}

%%%%%%%%%%%%%%%%%%%%%%%%%%%%%%%%%%%%%%%%%%%%%%%%%%%%%%%%%%%%%%%%%%%%%%%%%%%%%%%%%%%%%%%%%%%
\begin{figure}
\centerline{\psfig{width=5.5cm,angle=0,file=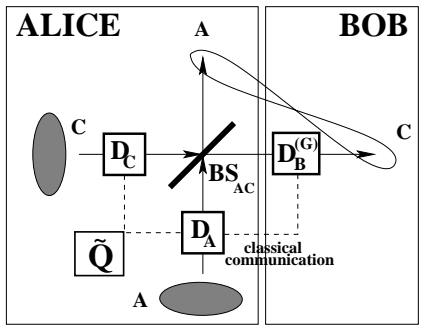}}
\caption{Scheme of the protocol for recovery of the entanglement between
modes $A$ and $C$. Mode $A$ in a suitable momentum-squeezed vacuum
state and mode $C$ in a position-squeezed vacuum state with the same level
of squeezing are displaced by displacements $D_{A}$ and $D_{C}$ distributed randomly
with Gaussian distribution characterized by the covariance matrix $\tilde{Q}$ (\ref{tildeQ}).
Next, the two modes are mixed on a balanced beam splitter $BS_{AC}$ that prepares
them in a two-mode separable state. Classical information traveling to Bob
is then used for displacement $D_{B}^{(G)}$ of mode $C$ with a suitable electronic gain $G$.
Provided that the gain is adjusted appropriately, the entanglement between modes $A$ and $C$ can be
perfectly recovered. See text for details.}
\label{fig3}
\end{figure}
%%%%%%%%%%%%%%%%%%%%%%%%%%%%%%%%%%%%%%%%%%%%%%%%%%%%%%%%%%%%%%%%%%%%%%%%%%%%%%%%%%%%%%%%%%%

The previous sections reveal that there indeed exist three-mode
mixed Gaussian fully separable states allowing to entangle two
modes of the state by mixing them stepwise with the third always
separable mode on two beam splitters. Therefore, once such a
tripartite state is established between distant Alice and Bob,
they can create, without need of any additional knowledge,
entanglement between them by sending a separable mode from one
site to another remote location. There is, however, another
interesting aspect in this protocol. When looking at the scheme in
Fig.~\ref{fig1}, one can imagine a different scenario when Bob has
some additional information about the preparation of the shared
state. Specifically, suppose that in each run of the protocol he
knows classical displacements $\bar{x}_{B},\bar{p}_{B}$ used to
displace his vacuum mode $B$. The natural question arises whether
Alice and Bob can then create a higher entanglement than when
using solely the shared fully separable state. In what follows we
give an affirmative answer to this question by showing that they
can in fact restore the whole amount of entanglement contained in
the two-mode squeezed vacuum state with CM $\gamma_{0,AC}$
defined in Eq.~(\ref{gammaACgammaB}). Moreover, here we provide an
explanation where the entanglement comes from in the improved protocol
depicted in Fig.~\ref{fig1}.

Let us consider the scheme in Fig.~\ref{fig3}. There, the
situation on Alice's side remains the same as in the protocol in
Fig.~\ref{fig1}. Bob, however, does not use an auxiliary vacuum
mode $B$ appropriately displaced with $D_{\rm B}$ to achieve the
desired final CM. Instead, he directly displaces mode $C$
received from Alice with a suitable electronic gain $G$. The CM of
the resulting state is most easily calculated in Heisenberg
picture. Let us denote by $x_{A}^{(0)}$, $p_{A}^{(0)}$,
$x_{C}^{(0)}$ and $p_{C}^{(0)}$ position and momentum quadratures
of initially vacuum modes $A$ and $C$, respectively. Arranging the
quadratures of the squeezed vacuum modes $A$ and $C$ into the
column vectors
$\xi_{A}=(x_{A},p_{A})^{T}=(e^{t}x_{A}^{(0)},e^{-t}p_{A}^{(0)})^{T}$
and
$\xi_{C}=(x_{C},p_{C})^{T}=(e^{-t}x_{C}^{(0)},e^{t}p_{C}^{(0)})^{T}$,
the quadratures of modes $A$ and $C$ after the displacements
$D_{A}$ and $D_{C}$ read $\xi_{A}+\bar{\xi}_{A}$ and
$\xi_{C}+\bar{\xi}_{C}$, where the vectors
$\bar{\xi}_{A}=(\bar{x}_{A},\bar{p}_{A})^{T}$ and
$\bar{\xi}_{C}=(\bar{x}_{C},\bar{p}_{C})^{T}$ define classical
displacements. Next, the two modes are mixed on the balanced beam
splitter $BS_{AC}$. Further, mode $C$ undergoes the displacement
$D_{B}^{(G)}$ defined by the displacement vector
$\bar{\xi}_{B}=(\bar{x}_{B},\bar{p}_{B})^{T}$ of mode $B$ and
$2\times2$ gain matrix $G$. This yields finally the transformed
quadratures in the following form:
%%%%%%%%%%%%%%%%%%%%%%%%%%%%%%%%%%%%%%%%%%%%%%%%%%%%%%%%%%%%%%%%%%%%%%%%%%%%%%%%%%%%%%%%%%%%%%%%%%%%%%%%
\begin{eqnarray}\label{transform}
\xi_{A}''&=&\frac{\xi_{A}+\bar{\xi}_{A}+\xi_{C}+\bar{\xi}_{C}}{\sqrt{2}},\nonumber\\
\xi_{C}''&=&\xi_{C}'+G\bar{\xi}_{B}=\frac{\xi_{A}+\bar{\xi}_{A}-\xi_{C}-\bar{\xi}_{C}}{\sqrt{2}}+G\bar{\xi}_{B}.\nonumber\\
\end{eqnarray}
%%%%%%%%%%%%%%%%%%%%%%%%%%%%%%%%%%%%%%%%%%%%%%%%%%%%%%%%%%%%%%%%%%%%%%%%%%%%%%%%%%%%%%%%%%%%%%%%%%%%%%%%
Here $\xi_{C}'$ is the vector of quadratures of mode $C$ behind
the beam splitter $BS_{AC}$. To determine the final state of a
two-mode subsystem $(AC)$, one has first to calculate the
correlations
$A_{ij}=\langle\{\xi_{A,i}'',\xi_{A,j}''\}\rangle$,
$C_{ij}=\langle\{\xi_{C,i}'',\xi_{C,j}''\}\rangle$ for
modes $A$ and $C$, respectively, and the intermodal correlations
$\Gamma_{ij}=\langle\{\xi_{A,i}'',\xi_{C,j}''\}\rangle$,
$i,j=1,2$, where the angle brackets denote averaging of the quantum quadrature
operators over the quantum state with CM
$\gamma_{A}\oplus\openone_{B}\oplus\gamma_{C}$,
where $\gamma_{A,C}$ are defined in Eq.~(\ref{product}) and the
classical displacements are averaged over the Gaussian distribution with covariance (\ref{tildeQ}).
Using Eqs.~(\ref{transform}), CMs (\ref{product}) and
the correlation matrix (\ref{tildeQ}) and taking into account the
fact that the vectors of quantum quadratures $\xi_{A},\xi_{C}$ are
uncorrelated with the vectors of classical displacements
$\bar{\xi}_{A},\bar{\xi}_{B}$ and $\bar{\xi}_{C}$, we arrive at
the CM of the state of modes $A$ and $C$
%%%%%%%%%%%%%%%%%%%%%%%%%%%%%%%%%%%%%%%%%%%%%%%%%%%%%%%%%%%%%%%%%%%%%%%%%%%%%%%%%%%%%%%%%%
\begin{eqnarray}\label{gamma2ACa}
\gamma_{2,AC}^{(G)}=\left(\begin{array}{cc}
A & \Gamma \\
\Gamma^{T} & C \\
\end{array}\right),
\end{eqnarray}
%%%%%%%%%%%%%%%%%%%%%%%%%%%%%%%%%%%%%%%%%%%%%%%%%%%%%%%%%%%%%%%%%%%%%%%%%%%%%%%%%%%%%%%%%%
where
%%%%%%%%%%%%%%%%%%%%%%%%%%%%%%%%%%%%%%%%%%%%%%%%%%%%%%%%%%%%%%%%%%%%%%%%%%%%%%%%%%%%%%%%%%
\begin{eqnarray}\label{tildeACGamma}
A&=&a\openone,\quad\Gamma=\sigma_{z}(b\openone+2xG^{T}),\nonumber\\
C&=&a\openone+4xGG^{T}-2x(G+G^{T}),
\end{eqnarray}
%%%%%%%%%%%%%%%%%%%%%%%%%%%%%%%%%%%%%%%%%%%%%%%%%%%%%%%%%%%%%%%%%%%%%%%%%%%%%%%%%%%%%%%%%%
where the parameters $a$ and $b$ are defined below Eq.(\ref{tildegamma2b}).
The structure of the CM indicates the optimal choice of the gain matrix $G$. We see that by choosing
$G=G_{\rm opt}=\openone$ the noise in the diagonal block $C$ cancels and the CM transforms to
%%%%%%%%%%%%%%%%%%%%%%%%%%%%%%%%%%%%%%%%%%%%%%%%%%%%%%%%%%%%%%%%%%%%%%%%%%%%%%%%%%%%%%%%%%
\begin{eqnarray}\label{gamma2ACb}
\gamma_{2,AC}^{(G_{\rm opt})}=\left(\begin{array}{cc}
a\openone & (b+2x)\sigma_{z} \\
(b+2x)\sigma_{z} & a\openone\\
\end{array}\right).
\end{eqnarray}
%%%%%%%%%%%%%%%%%%%%%%%%%%%%%%%%%%%%%%%%%%%%%%%%%%%%%%%%%%%%%%%%%%%%%%%%%%%%%%%%%%%%%%%%%%
To characterize the separability properties of CM
$\gamma_{2,AC}^{(G_{\rm opt})}$, it remains to calculate
the lower symplectic eigenvalue $\nu_{AC}$ for the matrix
$\left(\gamma_{2,AC}^{(G_{\rm opt})}\right)^{(T_{C})}$.
Using Eq.~(\ref{ABsymplectic}) one finds that it attains the
simple form $\nu_{AC}=a-b-2x=e^{-2t}$. Hence we
immediately see that the eigenvalue coincides with the lower
symplectic eigenvalue $\nu_{\rm in}=e^{-2t}$ of the
partially transposed pure two-mode squeezed vacuum state with CM
(\ref{gammaACgammaB}) that can be prepared by Alice from
her pure squeezed states. Since any Gaussian operation on mode $C$
cannot increase the entanglement of the state with CM (\ref{gamma2ACb})
\cite{Eisert_02} we see that Bob's displacement on mode $C$ with
the gain matrix $G_{\rm opt}$ is indeed an optimal Gaussian operation.
It is worth mentioning that while the entanglement of the two-mode squeezed
vacuum state figuring in the protocol can be restored perfectly,
the purity of the state cannot because
$\mbox{det}(\gamma_{2,AC}^{(G_{\rm
opt})})=\left(1+2xe^{-2t}\right)^{2}$.

Previous analysis reveals the origin of the entanglement in the
improved protocol in Fig.~\ref{fig1}. The displacements $D_{A}$ and
$D_{C}$ before the beam splitter $BS_{AC}$ can be transformed to
another displacements $D_{A}'$ and $D_{C}'$ behind the beam
splitter. Consequently, the state of modes $A$ and $C$ behind the
beam splitter can be viewed as a pure two-mode squeezed vacuum
state with CM (\ref{gammaACgammaB}). The entanglement and purity
of this state is then subsequently destroyed by local correlated
displacements $D_{A}'$ and $D_{C}'$. The classical information on
displacements $\bar{x}_{B}$ and $\bar{p}_{B}$ traveling to Bob,
that as we have seen allows him to perfectly restore the
entanglement between modes $A$ and $C$ if he uses the protocol of
Fig.~\ref{fig3}, is instead used to displace his vacuum mode $B$.
Mixing of the displaced vacuum mode $B$ with mode $C$ on the
second balanced beam splitter $BS_{BC}$ leads finally to the
following vector of Bob's quadratures:
%%%%%%%%%%%%%%%%%%%%%%%%%%%%%%%%%%%%%%%%%%%%%%%%%%%%%%%%%%%%%%%%%%%%%%%%%%%%%%%%%%%%%%%%%%%%%%%%%%%%%%%%
\begin{eqnarray}\label{transformb}
\xi_{B,{\rm out}}=\frac{\xi_{C}'+\bar{\xi}_{B}+\xi_{B}^{(0)}}{\sqrt{2}},
\end{eqnarray}
%%%%%%%%%%%%%%%%%%%%%%%%%%%%%%%%%%%%%%%%%%%%%%%%%%%%%%%%%%%%%%%%%%%%%%%%%%%%%%%%%%%%%%%%%%%%%%%%%%%%%%%%
where $\xi_{B}^{(0)}=(x_{B}^{(0)},p_{B}^{(0)})^{T}$ is the vector of Bob's vacuum quadratures and $\xi_{C}'$ denotes
the vector of quadratures of mode $C$ behind the beam splitter $BS_{AC}$. Compare the last equation with
the equation for the vector of optimal quadratures $\xi_{C}''=\xi_{C}'+\bar{\xi}_{B}$
obtained by setting $G=G_{\rm opt}=\openone$ in equation for $\xi_{C}''$ in Eqs.~(\ref{transform}).
We find out that the vector of output Bob's quadratures
(\ref{transformb}) is obtained from the vector of optimal quadratures by mixing it with the vector of vacuum
quadratures of Bob's mode on a balanced beam splitter described by Eq.~(\ref{UBC}).
Thus, mixing of mode $C$ with mode $B$ on a beam splitter $BS_{BC}$ in
Fig.~\ref{fig1} implements effectively optimal displacement of mode $C$ restoring perfectly entanglement between modes $A$ and $C$ followed
by mixing of the mode with the vacuum mode on a balanced beam splitter which reduces the amount of the entanglement. This finally
explains the origin of the distributed entanglement in the improved protocol as well as the fact that it is weaker in comparison
with the maximum entanglement that can be restored by Bob using the scheme in Fig.~\ref{fig3} with optimal gain matrix $G_{\rm opt}$.
%%%%%%%%%%%%%%%%%%%%%%%%%%%%%%%%%%%%%%%%%%%%%%%%%%%%%%%%%%%%%%%%%%%%%%%%%%%%%%%%%%%%%%%%%%
\section{Conclusion}\label{sec_5}
In conclusion, we performed a detailed investigation of the origin of entanglement in
the protocol for entanglement distribution by separable Gaussian states \cite{Mista_08}.
For this purpose we proposed a more simple and efficient protocol that enables to distribute
entanglement with more than an order of the magnitude higher logarithmic negativity than the older protocol
\cite{Mista_08}. Based on the new protocol we then found that the distributed entanglement
originates from the entanglement of sender's mode with the auxiliary mode that is used for distribution.
The entanglement is first destroyed by local correlated displacements that make the
auxiliary mode separable from sender's mode. The separable mode is then sent to the
receiver who partially restores the entanglement by mixing it with his suitably classically
correlated mode.

We emphasize that the proposed protocol for establishing entanglement between two remote
parties using separable ancilla mode consists of experimentally feasible Gaussian
states and operations involving single-mode squeezed states, correlated displacements and
beam splitters, dispensing with the technically challenging CNOT gates of the qubit case.
The input squeezed states with CMs (\ref{product}) need not to be minimum uncertainty states
and the protocol works well also with states with a large noise excess in the antisqueezed
quadratures, e.g., for $e^{2t}=2$ and $200$ vacuum units of noise excess in each quadrature the
logarithmic negativity of the distributed entanglement is $E_{\cal N}\approx0.5851$ ebits compared to
$E_{\cal N}\approx0.6019$ ebits for the minimum uncertainty states.

The verification of the effect, however, can become an issue, as
the possibility to demonstrate experimentally the proposed
protocol crucially relies on our ability to check separability of
mode $C$ from the pair of modes $(AB)$ in step 2. In theory we
assumed Gaussianity of states involved in the protocol. This
enabled us to prove the separability using relatively simple
sufficient conditions for separability of Gaussian states
\cite{Vidal_02,Serafini_06}. In an experiment, however, in order
to check whether the state prepared by us is Gaussian we would
have to perform complete tomography of the state on
infinite-dimensional state space which amounts to infinite number
of measurements. In this sense the Gaussian states are idealized
limit states and for the experimental demonstration of the
protocol we therefore need a sufficient condition for separability
of a generic generally non-Gaussian state or a reliable
indication, that the Gaussian character of the states is preserved
throughout the protocol. Nevertheless, a less ambitious goal can
be achieved by measuring only the three-mode CM of the state in
step 2. By showing that the measured CM satisfies some Gaussian
sufficient condition for separability of mode $C$ from modes
$(AB)$ we know that the Gaussian entanglement of modes $A$ and $B$
was distributed by sending an ancilla that was separable in
Gaussian sense.
%%%%%%%%%%%%%%%%%%%%%%%%%%%%%%%%%%%%%%%%%%%%%%%%%%%%%%%%%%%%%%%%%%%
\acknowledgments

L. M. would like to thank Radim Filip and Jarom\'{\i}r Fiur\'a\v{s}ek for fruitful discussions.
The research has been supported by the research projects ``Measurement and Information in Optics,'' (MSM 6198959213),
Center of Modern Optics (LC06007) of the Czech Ministry of Education and from GACR (Grant No. 202/08/0224).
We also acknowledge the financial support of the Future and Emerging Technologies (FET) programme
within the Seventh Framework Programme for Research of the European Commission, under the FET-Open grant agreement
COMPAS, number 212008. N. K. is grateful to the Alexander von Humboldt Foundation for the financial support.
%%%%%%%%%%%%%%%%%%%%%%%%%%%%%%%%%%%%%%%%%%%%%%%%%%%%%%%%%%%%%%%%%%
%%%%%%%%%%%%%%%%%%%%%%%%%%%%%%%%%%%%%%%%%%%%%%%%%%%%%%%%%%%%%%%%%%

%%%%%%%%%%%%%%%%%%%%%%%%%%%%%%%%%%%%%%%%%%%%%%%%%%%%%%%%%%%%%%%%%%%%%%%%%%%%
\end{document}